\documentclass[useAMS,usenatbib,twocolumn]{mnras}
\usepackage{graphicx}
\usepackage{dcolumn}   
\usepackage{bm}        
\usepackage[dvipsnames]{xcolor}
\usepackage{amssymb, amsmath,framed,mathtools}

\expandafter\ifx\csname package@font\endcsname\relax\else
 \expandafter\expandafter
 \expandafter\usepackage
 \expandafter\expandafter
 \expandafter{\csname package@font\endcsname}
\fi
\def\bq{\begin{equation}}
\def\eq{\end{equation}}
\def\bqy{\begin{eqnarray}}
\def\eqy{\end{eqnarray}}

\def\ga{\gamma}

\def\la{\lambda}

\begin{document}
\title[Artificial Spectral Edges]{Natural and Artificial Spectral Edges in Exoplanets}

\author[Lingam \& Loeb]{Manasvi Lingam$^{1,2}$\thanks{E-mail:
manasvi@seas.harvard.edu} and Abraham Loeb$^{2}$\thanks{E-mail:
aloeb@cfa.harvard.edu}\\
$^{1}$John A. Paulson School of Engineering and Applied Sciences, Harvard University, Cambridge, MA 02138, USA\\
$^{2}$Harvard-Smithsonian Center for Astrophysics, 60 Garden Street, Cambridge, MA 02138, USA}

\pagerange{\pageref{firstpage}--\pageref{lastpage}} \pubyear{2017}

\maketitle

\label{firstpage}

\begin{abstract}
Technological civilizations may rely upon large-scale photovoltaic arrays to harness energy from their host star. Photovoltaic materials, such as silicon, possess distinctive spectral features, including an ``artificial edge'' that is characteristically shifted in wavelength shortwards of the ``red edge'' of vegetation. Future observations of reflected light from exoplanets would be able to detect both natural and artificial edges photometrically, if a significant fraction of the planet's surface is covered by vegetation or photovoltaic arrays respectively. The stellar energy thus tapped can be utilized for terraforming activities by transferring heat and light from the day side to the night side on tidally locked exoplanets, thereby producing detectable artifacts.
\end{abstract}


\section{Introduction} \label{SecIntro}
In the rapidly advancing field of exoplanetary research, one of the most important areas is determining whether a given planet is capable of sustaining life \citep{Cock16}, and, if so, identifying the different avenues by which signatures of life may be detected \citep{SBP16}. It is widely accepted that life engenders chemical disequilibrium in planetary atmospheres \citep{Love65}. Thus, the most commonly studied method of detecting life entails the search for biosignature gases, such as oxygen and methane, occurring at disproportionately high levels \citep{SD10}. Owing to the predominance and smaller size of M-dwarfs, many studies have been conducted pertaining to the detection of biosignature gases from habitable planets transiting such stars \citep[e.g][]{Sc07}. In this context, the recent discovery of Proxima Centauri b \citep{AEs16}, the nearest exoplanet, represents an excellent opportunity to investigate the existence and composition of its atmosphere \citep{KL16}.

However, there are problematic issues with detecting biosignature gases, such as false positives arising from abiotic production \citep{Har15}. Hence, alternative means of detecting extraterrestrial life have also been pursued concomitantly. One of the most notable amongst them is the prospect of detecting vegetation by means of the ``red edge'' arising from a rapid change in the reflectance of vegetation around $0.7$ $\mu$m \citep{STSF,KSGB07}. The identification of the red edge can be viewed as a particular case of the more general statement that photometric observations of exoplanets can reveal many of their fundamental properties \citep{FST01}, such as the fractional coverage of their surface by oceans, land and vegetation \citep{MRP06,CAM09,FKS10}; the former may also be identifiable by means of the ``glint'' effect \citep{WG08}.

In this Letter, we study the potential existence, implications and detection of \emph{artificial} spectral edges on exoplanets, with a particular focus on tidally locked planets in the habitable zone (HZ) of M-dwarfs, such as Proxima b. 

\section{The case for artificial spectral edges}
We explore how and why artificial spectral edges could be operational on planets with surfaces modified by advanced civilizations. By an ``artificial edge'', we refer to a distinctive change in the reflectivity over a fairly narrow bandwidth, analogous to the red edge arising from vegetation.

The peak incident photon flux on planets orbiting M-dwarfs is shifted towards longer wavelengths. Hence, the ``red edge'' of vegetation on these planets may manifest in the near-IR region, possibly at wavelengths $\lesssim 1.1$ $\mu$m \citep{KST07}. Nevertheless, the EUV flux on the day side of Proxima b is $\sim 30$ times larger than on Earth \citep{Rib16}, potentially making spectral features in the EUV reflectance detectable as well.

\subsection{The case for solar energy}
On Earth, most sources of energy have been derived either directly or indirectly from solar power. To meet the growing energy needs, there has been a greater emphasis on renewable sources of energy, on account of its advantages \citep{JaDe11}. In particular, the production costs of solar energy have declined rapidly \citep{Baz13}. 

Many of these facts ought to be applicable to other advanced civilizations. One may therefore assume that they require energy for sustenance, and that it would be more sensible to rely on renewable sources such as stellar energy. We do not consider geothermal energy or nuclear fusion in this manuscript, since it is not apparent as to what kind of detectable signatures would be produced. Stellar energy is particularly relevant for habitable exoplanets around M-dwarfs, that are typically tidally locked \citep{Sc07}. The day side would receive continual illumination which enhances the effectiveness of exploiting stellar energy. Thus, we shall henceforth adopt the premise that advanced civilizations cover a significant fraction of the day side of the planet with photovoltaic devices (solar cells), which are robust, efficient and versatile \citep{MaCa03}.

A couple of caveats must be noted here. If the atmosphere is hazy, surface-based photovoltaic arrays would prove to be ineffective. In addition, tidally locked planets may also experience strong winds, which raise dust storms and lower the effectiveness of the photovoltaic arrays. An important factor to be borne in mind is the potential existence of a sub-solar cloud cover, which depends on the planetary rotation rate and other factors \citep{SBJ16}, may lower the effectiveness of such arrays, and also make the resultant signatures undetectable. 

Some of these limitations may be surmountable, in principle, by situating the arrays above the atmosphere (in space), or on bare rocky planets (with no atmosphere) that are used purely for harvesting stellar energy. In the latter scenario, the power would be beamed back to the inhabited world, and efficiencies $\gtrsim 50\%$ have been argued to be attainable \citep{Ben08}.

\subsection{The case for silicon} \label{SSecSil}
Currently, most solar cells are made out of pure silicon in different forms, which, despite its limitations \citep{Gr01}, is widely employed on account of its relative abundance. Although we focus on silicon henceforth, our analysis is valid for other photovoltaic materials that display an ``absorption edge'' \citep{Nelson03,MaCa03}.

Advanced civilizations may conceivably inhabit planets where photovoltaic materials with higher efficiencies are commonly available, implying that their counterparts of solar cells would not be silicon-based. Although this case cannot (and ought not) be dismissed, there are important reasons that favor the use of silicon for producing mega-scale photovoltaic arrays. 

The primary reason is the high cosmic abundance of silicon compared to other photovoltaic elements. This statement has considerable evidence, both empirical and theoretical, as heavy elements are generated via supernovae nucleosynthesis \citep{WH07,NKT13}. Hence, in purely probabilistic terms, planets are more likely to be endowed with larger reserves of silicon. Moreover, the importance of silicon in habitability, especially in the Galactic context, has also been documented \citep{GBW01}. The abundance of silicon is also closely linked with the cost of refining, processing and manufacturing silicon-based solar cells \citep{Baz13}. For instance, the present economic expenditures involved in producing gallium arsenide photovoltaics are approximately $2$-$3$ orders of magnitude higher than their silicon counterparts \citep{MaCa03}.

One other crucial reason is more viable for G-type (and perhaps K-type) stars, such as the Sun. The seminal work of \citet{SQ61} enabled the computation of the maximum theoretical efficiency of a solar cell (which uses one p-n junction). As per modern estimates, it is around $33\%$ for a band gap of around $E_\mathrm{max} = 1.34$ eV \citep{Ruh16}. Silicon has a band gap of $E_\mathrm{Si} = 1.1$ eV,\footnote{The band-gap for gallium arsenide is $1.42$ eV \citep{MaCa03}, while the corresponding value for perovskites is dependent on the chemical composition, although typically $\gtrsim 1.5$ eV in most instances \citep{GHBS14}.} which is close to $E_\mathrm{max}$. Another important factor considered by \citet{SQ61} is the ``ultimate efficiency'', which considers the efficiency only based on spectrum losses, ignoring other factors such as impedance matching. The ultimate efficiency is higher than the aforementioned theoretical efficiency, occurring at a band gap $E_g$ of
\begin{equation}
    E_g \approx 2.2 k_B T_s,
\end{equation}
where $k_B$ is the Boltzmann factor and $T_s$ is the blackbody temperature of the star. Using $T_s = 5770$ K for the Sun, we see that $E_g \approx 1.1$ eV, which equals $E_\mathrm{Si}$. Thus, silicon constitutes an effective photovoltaic material for G-type stars. 

For M-dwarfs, the efficiency of silicon, as per the basic estimates in \citet{SQ61}, will be lowered. But, the Shockley-Queisser limit can be bypassed in several ways, ranging from multiple p-n junctions to concentrating starlight \citep{Nelson03}. In addition, the existence of continual starlight on tidally locked planets, which occur more often around M-dwarfs, may also compensate for any reduction in the conversion efficiency. 

\subsection{Manifestation of the artificial edge} \label{SSecArtEdge}
\begin{figure}
\begin{center}
\includegraphics[width=7.5cm]{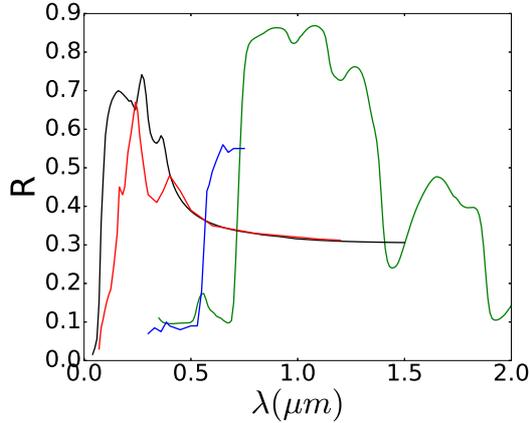}
\end{center}
\caption{The reflectance $R$ as a function of wavelength $\lambda$ for silicon (black), gallium arsenide (red), perovskite (blue) and vegetation (green) was adapted from \citet{green2008}, \citet{Blak82}, \citet{Ryu14} and \citet{clark2007} respectively.}
\label{Reflectance}
\end{figure}

One can therefore envision planets covered with mega-scale arrays of solar cells made of silicon. To quantify the chances of detecting them, the reflectance of silicon is plotted in Fig. \ref{Reflectance} based on Table 1 and Figure 1 of \citet{green2008}. The marked increase in the reflectivity of silicon in the UV region is nearly as dramatic as the red edge of vegetation on Earth at $0.7$ $\mu$m. Hence, the former is dubbed an ``artificial edge'' on account of its similarity with the latter. 

Although we have focused on silicon for reasons outlined earlier, artificial spectral edges would also be realized in the spectra of other photovoltaic materials \citep{Gr01} as well as artificial photosynthesis, for e.g., photocatalytic water splitting using titanium dioxide. This is explicitly illustrated in Fig. \ref{Reflectance} for gallium arsenide and organometal halide perovskites; the artificial edge for the latter is manifested at visible wavelengths $> 0.5$ $\mu$m \citep{Ryu14,GHBS14}, which may enhance the chances of detection. Hence, a significant change in the reflectance within a fairly narrow band that cannot be explained by natural causes (e.g. vegetation, oceans, soil) merits further investigation.

\begin{figure*}
\begin{center}
\includegraphics[width=14.8cm]{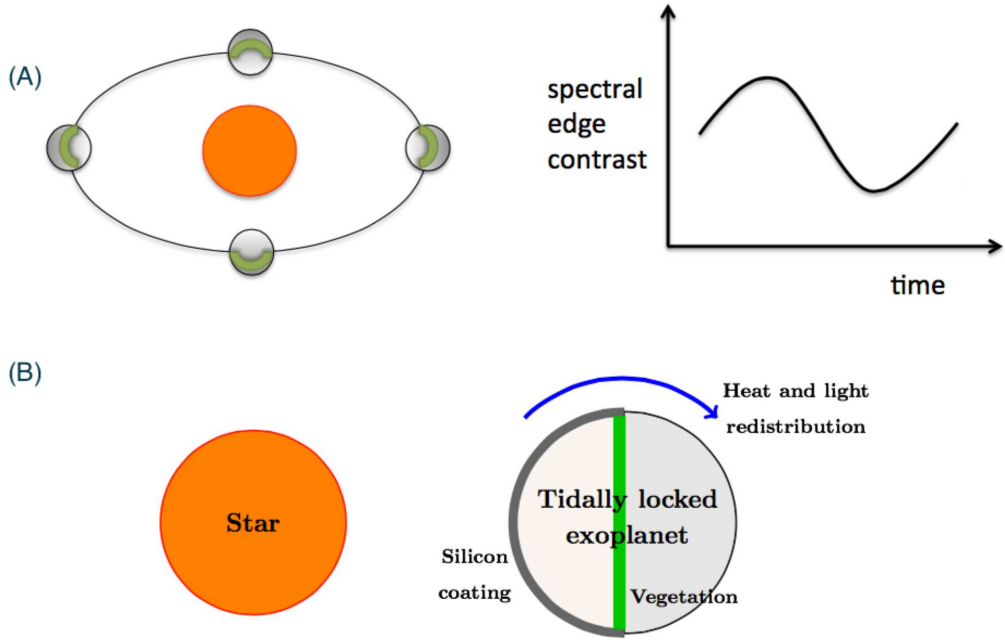}
\end{center}
\caption{(A) As the fraction of the planet covered by vegetation or photovoltaic arrays (silicon coating) changes during the orbit, so does the photometric flux in the reflected light just shortward of the wavelength characterizing their respective spectral edges. (B) Schematic illustration of terraforming on tidally locked exoplanets. Photovoltaic arrays on the day side are used to harness stellar energy, which is redistributed as heat and light on the night side. The green strip signifies the existence of vegetation close to the terminator.}
\label{Schematic}
\end{figure*}

Another striking similarity between natural vegetation and photovoltaic arrays is that, on tidally locked planets, they would most likely exist on the day side. The fraction of the exoplanet's surface covered by edge-producing materials can be deduced by analyzing the temporal photometric variability \citep{FST01}. A schematic illustration of this approach has been presented in Fig. \ref{Schematic}A. The basic idea is that, as the planet's orientation with respect to the observer changes, the fraction of the surface covered by photovoltaics would also be altered. In turn, this would lead to variations in the photometric flux, which can then be inverted to compute the fraction - a similar approach for oceans, land and vegetation was espoused in \citet{FKS10}.

\subsection{Detecting artificial edges}
The search for ``artificial edges'' on exoplanets is warranted and feasible with future telescopes, such as the Wide Field Infrared Survey Telescope (WFIRST), Large UV/Optical/Infrared Surveyor (LUVOIR), Habitable Exoplanet Imaging Mission (HabEx), and High-Definition Space Telescope (HDST). The change in the reflected-light contrast, i.e. the fraction of the starlight that is reflected from the planet's surface, is
\begin{equation}
    C = f R \left(\frac{R_p}{\sqrt{2} a}\right)^2 = 0.36\, \mathrm{ppm}\, f R \left(\frac{R_p}{R_\oplus}\right)^2 \left(\frac{a}{0.05\,\mathrm{AU}}\right)^{-2}
\end{equation}
where $R$ is the reflectance, $f$ is the fraction of the surface covered by photovoltaic panels, $R_p$ and $a$ are the radius of the planet and its distance from the star respectively. Since $C$ is a measurable quantity, which is dependent on the wavelength, one may expect to gain an estimate of $f \cdot R$. If a spectral edge is present, a noticeable change in $C$ should be manifest over a narrow range of wavelengths \citep{Winn10}. In turn, one can discern the fraction of the surface \citep{FKS10} covered by photovoltaics provided that its reflectivity is approximately known.

We have chosen to normalize $a$ in terms of Proxima b's semi-major axis \citep{AEs16}. The characteristic values of $C$ fall within the proposed sensitivity of WFIRST ($10^{-3}$ ppm) and LUVOIR ($10^{-4}$ ppm),\footnote{\noindent\url{https://asd.gsfc.nasa.gov/luvoir/docs/tech/Bolcar_SPIE_2015_paper.pdf}} even for relatively small values of $f$. By drawing upon the analogies with the red edge, a cloud-free coverage (by photovoltaic arrays) that is $\gtrsim 10\%$ of the planet's surface may suffice to make these large-scale structures detectable by these next-generation telescopes (Sec. 10 of \citealt{Sch17}).

The planet's atmosphere, if present, filters the reflected light from the surface, but the artificial edge could still be manifested, albeit in a modulated form.

\section{Implications of artificial edges}
We examine the possibility of detecting artificial edges either directly or indirectly, and the inferences that can be drawn from their existence.

In principle, one can infer the fractional area of vegetation through photometric observations \citep{FKS10}. Owing to the similarities between artificial and red edges discussed in Section \ref{SSecArtEdge}, $f$ could be computed. The stellar power captured is
\begin{equation} \label{PowPV}
P = \eta f \pi S \left(1-A\right) R_p^2,
\end{equation}
where $S$ and $A$ are the stellar irradiance and Bond albedo, and $\eta$ is the conversion efficiency of the photovoltaic cell. Thus, a measurement of $f$ in conjunction with (\ref{PowPV}) enables a heuristic estimation of the civilization class on the Kardashev scale \citep{Kard64}.

\subsection{Exploiting the generated power}

The power generated from these mega-scale photovoltaic arrays can be put to many uses. Solar cells can be particularly useful on tidally locked planets around M-dwarfs since the day side receives continual illumination. The extracted energy could power a plethora of processes on the night side:\\

\noindent{\bf (i) Heat redistribution:} On M-dwarfs, it may be possible that the day side is too hot for the sustenance of life (and photosynthesis), whilst the night side is too cold if atmosphere- and ocean-based redistribution is inefficient \citep{SBJ16}. It may thus be desirable to generate energy on the day side and redistribute it artificially to the night side for this purpose.

\noindent{\bf (ii) Light generation:} A requirement for photosynthesis is the availability of photons. The night side enshrouded in perpetual darkness may prove inimical to photosynthesis. Generating large-scale solar energy is advantageous, since it could be tapped for generating light on the night side to induce photosynthesis. The consequently larger population of autotrophs would be capable of supporting greater numbers of heterotrophs.\\

A schematic depiction of these terraforming processes is illustrated in Fig. \ref{Schematic}B. All of the above operations have important ramifications in the Search for Extraterrestrial Intelligence (SETI). If vegetation is discovered on the night side at levels that are higher than predictions based on conventional models, it can be indicative of artificial heat and light redistribution, thereby pointing towards the existence of an advanced civilization.

Detecting the existence and composition of exoplanetary atmospheres and oceans, and determining the surface temperature are both technically feasible \citep{FKS10}. Hence, if a discrepancy between the two is detected, i.e. evidence of heat redistribution that exceeds the theoretical predictions (based on the atmosphere and ocean circulation), this may serve as a signature of extraterrestrial technology reliant on stellar energy. This signature could be detected even with JWST, based on the analysis in \citet{KL16}. \citet{LoTu12} proposed techniques for detecting artificial illumination, albeit in the Solar system context. Anomalous lighting, \emph{primarily} on the night side of a tidally locked exoplanet, could signify the existence of extraterrestrial activity based on stellar energy redistribution. The feasibility of detecting artificial lighting on the dark side has been recently explored in \citet{KSLG} (see also \citealt{Arn05}), and it was concluded that it falls within the capabilities of JWST.

The above effects can also help differentiate between false positives such as minerals like enstatite \citep{HES12} and photovoltaic arrays, since only the latter would be accompanied by anomalous heat and light redistribution. These statements are \emph{specific} to stellar power, since other energy sources ought not be dependent on the distinction between the day and night sides. Thus, the existence of mega-scale photovoltaic arrays are also detectable through the above routes.

\subsection{Connections with Dysonian SETI}
Our basic premise of advanced civilizations constructing large-scale arrays falls under the purview of `Dysonian' SETI - looking for alien mega-structures \citep{BCD11,WCZ16}. But, we note that the civilizations considered here are typically of the Kardashev I type, unlike most Dysonian SETI projects that have investigated Kardashev II or III civilizations.

Alien civilizations may have become extinct, or moved on elsewhere, but left behind these arrays. Although such devices would undergo wear and tear, they can remain functional for a duration of time that is not insignificant by astrophysical standards, and would thus represent genuine extraterrestrial artifacts (albeit ``recent'' ones). Although interstellar archaeology, in general, lies beyond our current technological capabilities \citep{Carr12}, the scenario studied herein ought to constitute a genuine counterexample. Another such case is industrial pollution in the atmospheres of exoplanets, which could outlast the civilizations that produced it \citep{LGL14}.

We have based our discussion on habitable exoplanets, mostly those around M-dwarfs, but a wider array of possibilities exist. Large-scale photovoltaic arrays are capable of functioning on habitable exomoons \citep{Hell14}, although they are truly effective only when the incident stellar flux is significant. They may even be present on `uninhabitable' planets, such as those without an atmosphere or with high temperatures. These panels can also exist on artificial mega-structures, such as Dyson spheres, for the purpose of generating electrical energy. This energy could be used to power beacons or drive light sails, as recently hypothesized in \citet{LL17}.

\section{Conclusion}
In this Letter, we briefly discussed the detectability of vegetation via the spectral ``red edge'' and followed this up by positing the existence of ``artificial'' edges that can be studied along similar lines. 

Our hypothesis stemmed from two simple considerations. The first is that a civilization needs a power supply, which is easily met via stellar energy. We reasoned that mega-scale photovoltaic arrays for extracting stellar energy should be made of silicon on account of its wide availability, optimal processing and manufacturing, and reasonable efficiency. We highlighted the distinctive spectral features of silicon which could make the detection of such arrays feasible, provided that: (i) they are large enough, (ii) the viewing angles are favorable, and (iii) the cloud cover is limited.  

The implications arising from the existence of large-scale photovoltaic arrays were explored as well as the parent civilization's Kardashev class. We argued that the harnessed energy can be employed for undertaking a host of terraforming activities such as: (i) heat and light redistribution, (ii) induced photosynthesis on the night side of tidally locked planets, and (iii) generating artificial beams. All of these possibilities are likely to be detectable in their own right, thereby improving the chances of testing our hypothesis. In fact, (i) represents a better chance of detecting the artificial spectral edge indirectly, instead of finding them via direct photometric studies. The search for these artifacts belongs to the category of Dysonian SETI, which has attracted a great deal of attention in recent times. 

Although we focused primarily on silicon-based cells, the analysis is valid for other photovoltaics, photocatalysts that are widely utilized in artificial photosynthesis, fuels and raw materials (such as steel, carbon nanotubes and glass) used for constructing mega-structures. In all these examples, clear-cut patterns in the photometric variations might be manifested that are not explicable by natural features, thereby serving as indicators of advanced civilizations.

\section*{Acknowledgments}
We thank Freeman Dyson, James Guillochon, Nia Imara, Laura Kreidberg, Zac Manchester, Gary Melnick, Jill Tarter, Ed Turner and the referee for their helpful comments concerning the manuscript. ML is grateful to Chuanfei Dong for his generous assistance. This work was supported by a grant from the Breakthrough Prize Foundation for the Starshot Initiative.


\begin{thebibliography}{}
\makeatletter
\relax
\def\mn@urlcharsother{\let\do\@makeother \do\$\do\&\do\#\do\^\do\_\do\%\do\~}
\def\mn@doi{\begingroup\mn@urlcharsother \@ifnextchar [ {\mn@doi@}
  {\mn@doi@[]}}
\def\mn@doi@[#1]#2{\def\@tempa{#1}\ifx\@tempa\@empty \href
  {http://dx.doi.org/#2} {doi:#2}\else \href {http://dx.doi.org/#2} {#1}\fi
  \endgroup}
\def\mn@eprint#1#2{\mn@eprint@#1:#2::\@nil}
\def\mn@eprint@arXiv#1{\href {http://arxiv.org/abs/#1} {{\tt arXiv:#1}}}
\def\mn@eprint@dblp#1{\href {http://dblp.uni-trier.de/rec/bibtex/#1.xml}
  {dblp:#1}}
\def\mn@eprint@#1:#2:#3:#4\@nil{\def\@tempa {#1}\def\@tempb {#2}\def\@tempc
  {#3}\ifx \@tempc \@empty \let \@tempc \@tempb \let \@tempb \@tempa \fi \ifx
  \@tempb \@empty \def\@tempb {arXiv}\fi \@ifundefined
  {mn@eprint@\@tempb}{\@tempb:\@tempc}{\expandafter \expandafter \csname
  mn@eprint@\@tempb\endcsname \expandafter{\@tempc}}}

\bibitem[\protect\citeauthoryear{{Anglada-Escud{\'e}}
  et~al.,}{{Anglada-Escud{\'e}} et~al.}{2016}]{AEs16}
{Anglada-Escud{\'e}} G.,  et~al., 2016, \mn@doi [Nature] {10.1038/nature19106},
  \href {http://adsabs.harvard.edu/abs/2016Natur.536..437A} {536, 437}

\bibitem[\protect\citeauthoryear{{Arnold}}{{Arnold}}{2005}]{Arn05}
{Arnold} L.~F.~A.,  2005, \mn@doi [ApJ] {10.1086/430437}, \href
  {http://adsabs.harvard.edu/abs/2005ApJ...627..534A} {627, 534}

\bibitem[\protect\citeauthoryear{{Bazilian} et~al.,}{{Bazilian}
  et~al.}{2013}]{Baz13}
{Bazilian} M.,  et~al., 2013, \mn@doi [Renewable Energy]
  {10.1016/j.renene.2012.11.029}, 53, 329

\bibitem[\protect\citeauthoryear{{Benford}}{{Benford}}{2008}]{Ben08}
{Benford} J.,  2008, \mn@doi [ITPS] {10.1109/TPS.2008.923760}, \href
  {http://adsabs.harvard.edu/abs/2008ITPS...36..569B} {36, 569}

\bibitem[\protect\citeauthoryear{{Blakemore}}{{Blakemore}}{1982}]{Blak82}
{Blakemore} J.~S.,  1982, \mn@doi [JAP] {10.1063/1.331665}, \href
  {http://adsabs.harvard.edu/abs/1982JAP....53..123B} {53, R123}

\bibitem[\protect\citeauthoryear{{Bradbury}, {Cirkovic}  \&
  {Dvorsky}}{{Bradbury} et~al.}{2011}]{BCD11}
{Bradbury} R.~J.,  {Cirkovic} M.~M.,   {Dvorsky} G.,  2011, Journal of the
  British Interplanetary Society, \href
  {http://adsabs.harvard.edu/abs/2011JBIS...64..156B} {64, 156}

\bibitem[\protect\citeauthoryear{{Carrigan}}{{Carrigan}}{2012}]{Carr12}
{Carrigan} R.~A.,  2012, \mn@doi [AcAau] {10.1016/j.actaastro.2011.12.002},
  \href {http://adsabs.harvard.edu/abs/2012AcAau..78..121C} {78, 121}

\bibitem[\protect\citeauthoryear{{Clark}, {Swayze}, {Wise}, {Livo}, {Hoefen},
  {Kokaly}  \& {Sutley}}{{Clark} et~al.}{2007}]{clark2007}
{Clark} R.~N.,  {Swayze} G.~A.,  {Wise} R.,  {Livo} K.~E.,  {Hoefen} T.,
  {Kokaly} R.~F.,   {Sutley} S.~J.,  2007, USGS digital spectral library
  splib06a

\bibitem[\protect\citeauthoryear{{Cockell} et~al.,}{{Cockell}
  et~al.}{2016}]{Cock16}
{Cockell} C.~S.,  et~al., 2016, \mn@doi [AsBio] {10.1089/ast.2015.1295}, \href
  {http://adsabs.harvard.edu/abs/2016AsBio..16...89C} {16, 89}

\bibitem[\protect\citeauthoryear{{Cowan} et~al.,}{{Cowan} et~al.}{2009}]{CAM09}
{Cowan} N.~B.,  et~al., 2009, \mn@doi [ApJ] {10.1088/0004-637X/700/2/915},
  \href {http://adsabs.harvard.edu/abs/2009ApJ...700..915C} {700, 915}

\bibitem[\protect\citeauthoryear{{Ford}, {Seager}  \& {Turner}}{{Ford}
  et~al.}{2001}]{FST01}
{Ford} E.~B.,  {Seager} S.,   {Turner} E.~L.,  2001, \mn@doi [Nature]
  {10.1038/35091009}, \href {http://adsabs.harvard.edu/abs/2001Natur.412..885F}
  {412, 885}

\bibitem[\protect\citeauthoryear{{Fujii}, {Kawahara}, {Suto}, {Taruya},
  {Fukuda}, {Nakajima}  \& {Turner}}{{Fujii} et~al.}{2010}]{FKS10}
{Fujii} Y.,  {Kawahara} H.,  {Suto} Y.,  {Taruya} A.,  {Fukuda} S.,  {Nakajima}
  T.,   {Turner} E.~L.,  2010, \mn@doi [ApJ] {10.1088/0004-637X/715/2/866},
  \href {http://adsabs.harvard.edu/abs/2010ApJ...715..866F} {715, 866}

\bibitem[\protect\citeauthoryear{{Gonzalez}, {Brownlee}  \& {Ward}}{{Gonzalez}
  et~al.}{2001}]{GBW01}
{Gonzalez} G.,  {Brownlee} D.,   {Ward} P.,  2001, \mn@doi [Icarus]
  {10.1006/icar.2001.6617}, \href
  {http://adsabs.harvard.edu/abs/2001Icar..152..185G} {152, 185}

\bibitem[\protect\citeauthoryear{{Gr{\"a}tzel}}{{Gr{\"a}tzel}}{2001}]{Gr01}
{Gr{\"a}tzel} M.,  2001, \mn@doi [Nature] {10.1038/35104607}, \href
  {http://adsabs.harvard.edu/abs/2001Natur.414..338G} {414, 338}

\bibitem[\protect\citeauthoryear{{Green}}{{Green}}{2008}]{green2008}
{Green} M.~A.,  2008, \mn@doi [Solar Energy Materials and Solar Cells]
  {10.1016/j.solmat.2008.06.009}, 92, 1305

\bibitem[\protect\citeauthoryear{{Green}, {Ho-Baillie}  \& {Snaith}}{{Green}
  et~al.}{2014}]{GHBS14}
{Green} M.~A.,  {Ho-Baillie} A.,   {Snaith} H.~J.,  2014, \mn@doi [NaPho]
  {10.1038/nphoton.2014.134}, \href
  {http://adsabs.harvard.edu/abs/2014NaPho...8..506G} {8, 506}

\bibitem[\protect\citeauthoryear{{Harman}, {Schwieterman}, {Schottelkotte}  \&
  {Kasting}}{{Harman} et~al.}{2015}]{Har15}
{Harman} C.~E.,  {Schwieterman} E.~W.,  {Schottelkotte} J.~C.,   {Kasting}
  J.~F.,  2015, \mn@doi [ApJ] {10.1088/0004-637X/812/2/137}, \href
  {http://adsabs.harvard.edu/abs/2015ApJ...812..137H} {812, 137}

\bibitem[\protect\citeauthoryear{{Heller} et~al.,}{{Heller}
  et~al.}{2014}]{Hell14}
{Heller} R.,  et~al., 2014, \mn@doi [AsBio] {10.1089/ast.2014.1147}, \href
  {http://adsabs.harvard.edu/abs/2014AsBio..14..798H} {14, 798}

\bibitem[\protect\citeauthoryear{{Hu}, {Ehlmann}  \& {Seager}}{{Hu}
  et~al.}{2012}]{HES12}
{Hu} R.,  {Ehlmann} B.~L.,   {Seager} S.,  2012, \mn@doi [ApJ]
  {10.1088/0004-637X/752/1/7}, \href
  {http://adsabs.harvard.edu/abs/2012ApJ...752....7H} {752, 7}

\bibitem[\protect\citeauthoryear{{Jacobson} \& {Delucchi}}{{Jacobson} \&
  {Delucchi}}{2011}]{JaDe11}
{Jacobson} M.~Z.,  {Delucchi} M.~A.,  2011, \mn@doi [Energy Policy]
  {10.1016/j.enpol.2010.11.040}, 39, 1154

\bibitem[\protect\citeauthoryear{{Kardashev}}{{Kardashev}}{1964}]{Kard64}
{Kardashev} N.~S.,  1964, SvA, \href
  {http://adsabs.harvard.edu/abs/1964SvA.....8..217K} {8, 217}

\bibitem[\protect\citeauthoryear{{Kiang}, {Siefert}, {Govindjee}  \&
  {Blankenship}}{{Kiang} et~al.}{2007a}]{KSGB07}
{Kiang} N.~Y.,  {Siefert} J.,  {Govindjee}  {Blankenship} R.~E.,  2007a,
  \mn@doi [AsBio] {10.1089/ast.2006.0105}, \href
  {http://adsabs.harvard.edu/abs/2007AsBio...7..222K} {7, 222}

\bibitem[\protect\citeauthoryear{{Kiang} et~al.,}{{Kiang}
  et~al.}{2007b}]{KST07}
{Kiang} N.~Y.,  et~al., 2007b, \mn@doi [AsBio] {10.1089/ast.2006.0108}, \href
  {http://adsabs.harvard.edu/abs/2007AsBio...7..252K} {7, 252}

\bibitem[\protect\citeauthoryear{{Korpela}, {Sallmen}  \& {Leystra
  Greene}}{{Korpela} et~al.}{2015}]{KSLG}
{Korpela} E.~J.,  {Sallmen} S.~M.,   {Leystra Greene} D.,  2015, \mn@doi [ApJ]
  {10.1088/0004-637X/809/2/139}, \href
  {http://adsabs.harvard.edu/abs/2015ApJ...809..139K} {809, 139}

\bibitem[\protect\citeauthoryear{{Kreidberg} \& {Loeb}}{{Kreidberg} \&
  {Loeb}}{2016}]{KL16}
{Kreidberg} L.,  {Loeb} A.,  2016, \mn@doi [ApJL]
  {10.3847/2041-8205/832/1/L12}, \href
  {http://adsabs.harvard.edu/abs/2016ApJ...832L..12K} {832, L12}

\bibitem[\protect\citeauthoryear{{Lin}, {Gonzalez Abad}  \& {Loeb}}{{Lin}
  et~al.}{2014}]{LGL14}
{Lin} H.~W.,  {Gonzalez Abad} G.,   {Loeb} A.,  2014, \mn@doi [ApJL]
  {10.1088/2041-8205/792/1/L7}, \href
  {http://adsabs.harvard.edu/abs/2014ApJ...792L...7L} {792, L7}

\bibitem[\protect\citeauthoryear{{Lingam} \& {Loeb}}{{Lingam} \&
  {Loeb}}{2017}]{LL17}
{Lingam} M.,  {Loeb} A.,  2017, \mn@doi [ApJL] {10.3847/2041-8213/aa633e},
  \href {http://adsabs.harvard.edu/abs/2017ApJ...837L..23L} {837, L23}

\bibitem[\protect\citeauthoryear{{Loeb} \& {Turner}}{{Loeb} \&
  {Turner}}{2012}]{LoTu12}
{Loeb} A.,  {Turner} E.~L.,  2012, \mn@doi [AsBio] {10.1089/ast.2011.0758},
  \href {http://adsabs.harvard.edu/abs/2012AsBio..12..290L} {12, 290}

\bibitem[\protect\citeauthoryear{{Lovelock}}{{Lovelock}}{1965}]{Love65}
{Lovelock} J.~E.,  1965, \mn@doi [Nature] {10.1038/207568a0}, \href
  {http://adsabs.harvard.edu/abs/1965Natur.207..568L} {207, 568}

\bibitem[\protect\citeauthoryear{{Markvart} \& {Casta{\~n}er}}{{Markvart} \&
  {Casta{\~n}er}}{2003}]{MaCa03}
{Markvart} T.,  {Casta{\~n}er} L.,  2003, Practical Handbook of Photovoltaics:
  Fundamentals and Applications.
Elsevier: Oxford, U.K.

\bibitem[\protect\citeauthoryear{{Monta{\~n}{\'e}s-Rodr{\'{\i}}guez},
  {Pall{\'e}}, {Goode}  \&
  {Mart{\'{\i}}n-Torres}}{{Monta{\~n}{\'e}s-Rodr{\'{\i}}guez}
  et~al.}{2006}]{MRP06}
{Monta{\~n}{\'e}s-Rodr{\'{\i}}guez} P.,  {Pall{\'e}} E.,  {Goode} P.~R.,
  {Mart{\'{\i}}n-Torres} F.~J.,  2006, \mn@doi [ApJ] {10.1086/507694}, \href
  {http://adsabs.harvard.edu/abs/2006ApJ...651..544M} {651, 544}

\bibitem[\protect\citeauthoryear{{Nelson}}{{Nelson}}{2003}]{Nelson03}
{Nelson} J.,  2003, The Physics of Solar Cells.
World Scientific

\bibitem[\protect\citeauthoryear{{Nomoto}, {Kobayashi}  \& {Tominaga}}{{Nomoto}
  et~al.}{2013}]{NKT13}
{Nomoto} K.,  {Kobayashi} C.,   {Tominaga} N.,  2013, \mn@doi [ARA\&A]
  {10.1146/annurev-astro-082812-140956}, \href
  {http://adsabs.harvard.edu/abs/2013ARA%26A..51..457N} {51, 457}

\bibitem[\protect\citeauthoryear{{Ribas} et~al.,}{{Ribas} et~al.}{2016}]{Rib16}
{Ribas} I.,  et~al., 2016, \mn@doi [A\&A] {10.1051/0004-6361/201629576}, \href
  {http://adsabs.harvard.edu/abs/2016A%26A...596A.111R} {596, A111}

\bibitem[\protect\citeauthoryear{{R{\"u}hle}}{{R{\"u}hle}}{2016}]{Ruh16}
{R{\"u}hle} S.,  2016, \mn@doi [Solar Energy] {10.1016/j.solener.2016.02.015},
  \href {http://adsabs.harvard.edu/abs/2016SoEn..130..139R} {130, 139}

\bibitem[\protect\citeauthoryear{{Ryu}, {Noh}, {Jeon}, {Kim}, {Yang}, {Seo}  \&
  {Seok}}{{Ryu} et~al.}{2014}]{Ryu14}
{Ryu} S.,  {Noh} J.~H.,  {Jeon} N.~J.,  {Kim} Y.~C.,  {Yang} W.~S.,  {Seo} J.,
   {Seok} S.~I.,  2014, \mn@doi [Energy \& Environmental Science]
  {10.1039/C4EE00762J}, 7, 2614

\bibitem[\protect\citeauthoryear{{Scalo} et~al.,}{{Scalo} et~al.}{2007}]{Sc07}
{Scalo} J.,  et~al., 2007, \mn@doi [AsBio] {10.1089/ast.2006.0125}, \href
  {http://adsabs.harvard.edu/abs/2007AsBio...7...85S} {7, 85}

\bibitem[\protect\citeauthoryear{{Schwieterman} et~al.,}{{Schwieterman}
  et~al.}{2017}]{Sch17}
{Schwieterman} E.~W.,  et~al., 2017, submitted to AsBio (arXiv:1705.05791), \href
  {http://adsabs.harvard.edu/abs/2017arXiv170505791S} {}

\bibitem[\protect\citeauthoryear{{Seager} \& {Deming}}{{Seager} \&
  {Deming}}{2010}]{SD10}
{Seager} S.,  {Deming} D.,  2010, \mn@doi [ARA\&A]
  {10.1146/annurev-astro-081309-130837}, \href
  {http://adsabs.harvard.edu/abs/2010ARA%26A..48..631S} {48, 631}

\bibitem[\protect\citeauthoryear{{Seager}, {Turner}, {Schafer}  \&
  {Ford}}{{Seager} et~al.}{2005}]{STSF}
{Seager} S.,  {Turner} E.~L.,  {Schafer} J.,   {Ford} E.~B.,  2005, \mn@doi
  [AsBio] {10.1089/ast.2005.5.372}, \href
  {http://adsabs.harvard.edu/abs/2005AsBio...5..372S} {5, 372}

\bibitem[\protect\citeauthoryear{{Seager}, {Bains}  \& {Petkowski}}{{Seager}
  et~al.}{2016}]{SBP16}
{Seager} S.,  {Bains} W.,   {Petkowski} J.~J.,  2016, \mn@doi [AsBio]
  {10.1089/ast.2015.1404}, \href
  {http://adsabs.harvard.edu/abs/2016AsBio..16..465S} {16, 465}

\bibitem[\protect\citeauthoryear{{Shields}, {Ballard}  \& {Johnson}}{{Shields}
  et~al.}{2016}]{SBJ16}
{Shields} A.~L.,  {Ballard} S.,   {Johnson} J.~A.,  2016, \mn@doi [PhR]
  {10.1016/j.physrep.2016.10.003}, \href
  {http://adsabs.harvard.edu/abs/2016arXiv161005765S} {663, 1}

\bibitem[\protect\citeauthoryear{{Shockley} \& {Queisser}}{{Shockley} \&
  {Queisser}}{1961}]{SQ61}
{Shockley} W.,  {Queisser} H.~J.,  1961, \mn@doi [JAP] {10.1063/1.1736034},
  \href {http://adsabs.harvard.edu/abs/1961JAP....32..510S} {32, 510}

\bibitem[\protect\citeauthoryear{{Williams} \& {Gaidos}}{{Williams} \&
  {Gaidos}}{2008}]{WG08}
{Williams} D.~M.,  {Gaidos} E.,  2008, \mn@doi [Icarus]
  {10.1016/j.icarus.2008.01.002}, \href
  {http://adsabs.harvard.edu/abs/2008Icar..195..927W} {195, 927}

\bibitem[\protect\citeauthoryear{{Winn}}{{Winn}}{2010}]{Winn10}
{Winn} J.~N.,  2010, in {Seager} S.,  ed., Exoplanets. University of Arizona
  Press, pp 55--77

\bibitem[\protect\citeauthoryear{{Woosley} \& {Heger}}{{Woosley} \&
  {Heger}}{2007}]{WH07}
{Woosley} S.~E.,  {Heger} A.,  2007, \mn@doi [PhR]
  {10.1016/j.physrep.2007.02.009}, \href
  {http://adsabs.harvard.edu/abs/2007PhR...442..269W} {442, 269}

\bibitem[\protect\citeauthoryear{{Wright}, {Cartier}, {Zhao}, {Jontof-Hutter}
  \& {Ford}}{{Wright} et~al.}{2016}]{WCZ16}
{Wright} J.~T.,  {Cartier} K.~M.~S.,  {Zhao} M.,  {Jontof-Hutter} D.,   {Ford}
  E.~B.,  2016, \mn@doi [ApJ] {10.3847/0004-637X/816/1/17}, \href
  {http://adsabs.harvard.edu/abs/2016ApJ...816...17W} {816, 17}

\makeatother
\end{thebibliography}

\label{lastpage}

\end{document}